\documentclass[aps,pre,reprint,groupedaddress,showpacs]{revtex4-1}
\usepackage{amssymb}
\usepackage{latexsym}
\usepackage{amsfonts}
\usepackage{graphicx}
\usepackage{float}
\usepackage{amsmath}
\usepackage{mathcomp}
\usepackage{epsfig}
\usepackage{color}
\usepackage{hyperref}
\begin{document}
\title{Collective dynamics on a two-lane asymmetrically coupled TASEP with mutually interactive Langmuir Kinetics}
\author{Arvind Kumar Gupta}
\email[]{akgupta@iitrpr.ac.in}
\affiliation{Department of Mathematics, Indian Institute of Technology Ropar, Rupnagar-140001, Punjab, India.}

\begin{abstract}
Motivated by the recent experimental observations on clustering of motor proteins on microtubule filament, we study an open system of two parallel totally asymmetric simple exclusion processes under asymmetric coupling conditions, which incorporates the mutual interaction with the surrounding environment through Langmuir Kinetics in both the lanes. In the modified Langmuir Kinetics, the attachment and detachment rates depends on the configuration of nearest neighboring sites. We analyse the model within the framework of continuum mean-field theory and the phase diagrams along with density profiles are obtained using boundary layer analysis. The effect of mutual interactions on the phase diagram for two different situations of attachment and detachment (LK) rates is discussed. Under the symmetric LK dynamics, the topological structure of the phase diagram remains similar to the one in without mutual interaction; while for the antisymmetric case, after a certain critical value of attractive/repulsive mutual attraction, significant changes are found in the qualitative nature of phase diagram. Moreover, it is shown that the type of mutual interaction affects the dynamic properties of motor proteins. The theoretical findings are examined by extensive Monte-Carlo simulations.
\end{abstract}

%% insert suggested PACS numbers in braces on next line
%\keywords{Phase diagram, Two-lane, Langmuir Kinetics, Asymmetric coupling, Boundary layer analysis}
\maketitle
\section{\label{Intro}Introduction}
Many natural systems exhibit complex behavior under stationary state when either driven by some external field or self driven. Such driven diffusive systems reveal very rich nonequilibrium phenomena in physics, chemistry and biology such as kinetics of bio-polymerization \cite{macdonald1968kinetics},
dynamics of motor proteins in biological cells \cite{frey2004collective,chowdhury2013stochastic}, gel electrophoresis \cite{toroczkai1996model},
vehicular traffic \cite{sopasakis2006stochastic}, and modeling of ant-trails \cite{chowdhury2000statistical}. In biological systems, molecular motors are motor proteins that consume chemical energy and move along polymer filaments of the cytoskeleton, which act as macromolecular tracks. Their collective dynamics plays a major role in various intracellular processes/functions such as cellular trafficking, protein synthesis, cell division etc \cite{neri2013exclusion,frey2004collective,kolomeisky2013motor}.

In order to analyze the collective properties of interacting molecular motors, totally asymmetrically simple exclusion process (TASEP) model is found to be a paradigmatic model to study motion of motor proteins in the last decade \cite{krug1991boundary,parmeggiani2003phase,privman2005nonequilibrium}. TASEP model comprises of single species of particles performing biased hopping with uniform rate in a preferred direction along a 1D lattice. The particles obey certain preassigned rules under hard-core exclusion principle, due to which a lattice site cannot have more than one particle. In an open TASEP connected to boundary reservoirs, the particles remain conserved in the bulk. The conservation of particle number is violated by allowing particle adsorption (desorption) to (from) the bulk with fixed rates [Langmuir kinetics (LK)] similar to the attachment-detachment of motors occur form the filament~\cite{howardsinauer}. Single-channel TASEP coupled with LK has been well studied \cite{evans2003shock,parmeggiani2004totally,mirin2003effect} and the phase behavior is considerably different from those known in reference models of TASEP without LK \cite{krug1991boundary}. The competing dynamics of particle conservation (TASEP) and particle nonconservation (LK) results into distinguishing characteristics such as localization of shocks and phase coexistence \cite{evans2003shock,parmeggiani2003phase}. The bulk and surface transitions in a single TASEP with LK are examined via boundary layer analysis \cite{mukherji2006bulk}.

Experimental studies on collective transport of kinesin-1 motor protein along microtubule suggests that the individual motors interact with each other via short range interaction and this attraction is weakly attractive \cite{roos2008dynamic,vilfan2001dynamics,seitz2006processive}. The simulation results of the motor attachment/detachment dynamics in the presence of mutual interactions are found compatible with in-vitro experimental observations \cite{roos2008dynamic}. Moreover, this mutual interaction manifest as a propensity of the motors to cluster during their transport on microtubule. As a result, intermolecular interactions affect various chemical transitions inside the cell and play an important role in the collective dynamics of molecular motors. The results in refs. \cite{celis2015correlations,teimouri2015theoretical} further strengthen the fact that mutual interactions between motors plays a significant role in determining the spatial organization of the motors on filaments. To qualitatively understand the collective transport of weakly interactive molecular motors through short-range interactions on single filament, two different approaches namely mean field and modified cluster mean field approach is employed on TASEP model with \cite{chandel2015collective} and without LK \cite{celis2015correlations}, respectively. Recently, the effect of mutual interactions is examined in a one dimensional TASEP model by modifying the attcahment/detachment rates while keeping the hopping rate unaffected \cite{vuijk2015driven}. This is in contrast to the Katz-Lebowits-Spohn (KLS) model \cite{katz1984nonequilibrium} which include both the correlations due to mutual interactions and hard core exclusion principle.

It is evident that more realistic description of many transport processes such as motor protein, vehicular traffic etc. can be mimicked via multi-channel TASEPs. Moreover, kinesin (a motor-protein) can move along parallel protofilaments \cite{howard2002mechanics} forming a multi-lane system. Due to its importance, the two-channel system is studied extensively without LK under different coupling environments \cite{pronina2004two,pronina2006asymmetric,jiang2008weak}. Recently, the competition between bulk and boundary dynamics in steady state is examined by Gupta and Dhiman~\cite{gupta2014asymmetric} in a two-channel TASEP model with Langmuir kinetics in both the lanes under biased lane-changing rule. But the important aspect of mutual interaction is ignored in this study.

In this paper, we explore the consequences of mutually interacted LK on the stationary density profiles as well as on phase diagrams of a two-channel open TASEP under the fully asymmetric coupling environment. The proposed framework not only provide natural means to consider the interaction between the motors with the surrounding environment at micro level but also be useful to understand some aspects of the collective behavior of motor proteins that emerges at macro level in a coupled system. We also investigate the role of symmetry of interaction via LK rates and show that it significantly affect the system dynamic. The paper is organized as follows: In the next section, we firstly present the theoretical description of the two-lane TASEP model under the biased coupling environment. To incorporate the mutual interactions with the environment, the modified version of LK is proposed followed by the mean field approximation. In sec III, the method to obtain stationary density profiles is discussed. The phase diagram and density profiles under two different cases of modified LK rates are investigated in sec. IV. Finally, in sec. V, we provide our conclusions.

\section{Two-lane model and hydrodynamic mean-field approximation}\label{sec:1}
\begin{figure*}
\includegraphics[width=13cm,height=5.5cm]{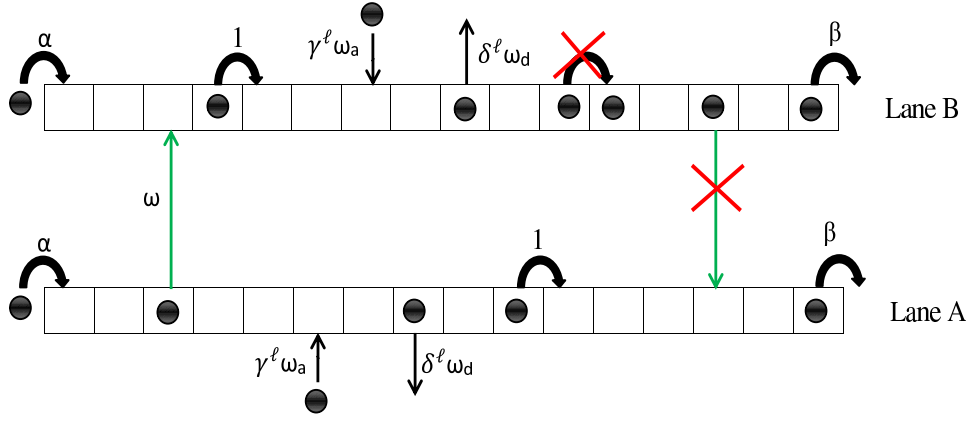}
\caption{\label{fig:1} Schematic diagram of the model. Crossed arrows indicate the forbidden transitions}
\end{figure*}

We define a two-lane open system of two parallel one-dimensional lattice channels, each with $N$ sites,
denoted by $A$ and $B$. The system consists of indistinguishable particles distributed under hard-core exclusion principle
 and hop unidirectionally to the right (See Fig.~\ref{fig:1}). The system configurations are characterized in terms of occupation numbers $
\tau_{i,j}$ ($i=1,2,3,.....N$; $j= A,B$), each of which is either zero (vacant site) or one
(occupied site). In an infinitesimal time interval $dt$, a lattice site $(i,j)$; $i=1,2,3,.....N$; $j= A,B$
is randomly chosen and the paticle hoping takes place according to the following dynamical rules.
\begin{enumerate}
\item If $i = 1$ (entrance of lane $j$), particles are injected into the lattice with
a rate $\alpha$ provided $\tau_{1,j} = 0$. If $\tau_{1,j} = 1$ and $\tau_{2,j} = 0$, then
 the particle moves from site $(1,j)$ to site $(2,j)$ with a unit rate. If $\tau_{1,j} = 1$
and $\tau_{2,j} = 1$, no hopping occurs from site $(1,j)$.~Neither lane changing nor
attachment-detachment takes place at $i = 1$.
\item If $i = N$ (exit of lane $j$), a particle is removed from the lattice with a rate $\beta$
when $\tau_{N,j}=1$. Neither lane changing nor attachment-detachment takes place at $i = N$.
\item For $1 < i < N$ (bulk of lane $j$), if $\tau_{i,A} = 0$, a particle can attach to the site
with a rate $\gamma^\ell \omega_a$ (attachment rate). When $\tau_{i,A} = 1$, then the particle at the site $(i, A)$ firstly tries to detach itself from the system with a rate $\delta^\ell \omega_d$ (detachment rate) and if it fails, then it moves forward to site $(i+1, A)$ with unit rate provided $\tau_{i+1,A} = 0$. If a particle at site $(i,A)$ fails to detach from the system followed by an unsuccessful attempt of moving forward, then it will try to shift to lane $B$ with a rate $\omega$; provided $\tau_{i,B} = 0$. Here the parameter $\ell$ depends on the configuration of the neighboring sites and is a non-negative integer which represents the strength of the modifying factors of attachment/detachment rates. The dynamics in the bulk of lane B are similar, with the only exception being that particles are forbidden to shift from lane B to lane A. Here, no lane-changing is allowed from lane $B$ to lane $A$.
\end{enumerate}
Here the mutual interactions are incorporated by modifying the Langmuir Kinetics in terms of detachment (attachment) rate from (to) the background. The modified LK dynamics at site $(i, j)$ depend on the configuration of two nearest neighboring sites according to the following rules:
\begin{itemize}
  \item If both the neighboring sites are vacant i.e. $\tau_{i-1,j}=0$ and $\tau_{i+1,j}=0$, then the parameter $\ell =0$ resulting the attachment (detachment) rate as $w_a$ ($w_d$).
  \item If either of the neighboring sites are vacant i.e. $\tau_{i-1,j}=1$ and $\tau_{i+1,j}=0$; or $\tau_{i-1,j}=0$ and $\tau_{i+1,j}=1$, $\ell$ takes the value 1 and the attachment (detachment) rate becomes $\gamma w_a$ ($\delta w_d$).
  \item If none of the neighboring sites are vacant i.e. $\tau_{i-1,j}=1$ and $\tau_{i+1,j}=1$, attachment (detachment) rate adapted as $\gamma^2 w_a$ ($\delta^2 w_d$) due to $\ell =2$.
\end{itemize}
%
%\begin{itemize}
%  \item $0 1 0 \longrightarrow 0 0 0$ with rate $w_d$.
%  \item $1 1 0 \longrightarrow 1 0 0$ with rate $qw_d$.
%  \item $0 1 1 \longrightarrow 0 0 1$ with rate $qw_d$.
%  \item $1 1 1 \longrightarrow 1 0 1$ with rate $q^2 w_d$.
%\end{itemize}

% At entrance ($i=1$), a particles are injected into the lattice with a rate $\alpha$
%provided $\tau_{1,j}=0$; and at exit ($i=N$), a particles are removed from the lattice with a rate $\beta$
%when $\tau_{N,j}=1$.
%moves forward to site $(i+1,A)$ with unit rate provided $\tau_{i+1,A}=0$; otherwise it shifts to lane B with a rate $w$ only if $\tau_{i,B}=0$. In the bulk, if $\tau_{i,A}=0$, a particle attachs to the site with a rate $\delta^\ell w_a$ (attachment rate).
For $\gamma=\delta=1$, the proposed model reduces to a two-lane asymmetrically coupled TASEP with LK in both the lanes \cite{gupta2014asymmetric}.
The rules for modified LK rates in both the lanes imparts the
model its generality over the existing two-lane coupled TASEP models with LK
\cite{jiang2007two,gupta2014asymmetric}. Note that our model
is suitable to study a number of two-lane transport processes such as vehicular traffic
and motor proteins because normally a vehicle or a molecular motor does not change
its lane unless hindered by another one preceding it. This is completely in accordance
with the lane-changing rules defined by us.

The master equations for the temporal evolution of occupancy of particles at each bulk site ($1<i<N$) in both the lanes ($j=A, B$)
 is given by
\begin{equation}
\begin{split}
\frac{d\textlangle \tau_{i,j}\textrangle}{dt}=\textlangle \tau_{i-1,j}(1-\tau_{i,j})\textrangle - \textlangle \tau_{i,j}(1-\tau_{i+1,j})\textrangle \\+ \omega_a \textlangle (1-\tau_{i-1,j})(1-\tau_{i,j})(1-\tau_{i+1,j})\textrangle \\+ \gamma\omega_a \textlangle (1-\tau_{i,j})[(1-\tau_{i+1,j})\tau_{i-1,j}+(1-\tau_{i-1,j})\tau_{i+1,j}]\textrangle \\+\gamma^2\omega_a \textlangle \tau_{i-1,j}(1-\tau_{i,j})\tau_{i+1,j} \textrangle \\-\omega_d\textlangle \tau_{i,j}(1-\tau_{i-1,j})(1-\tau_{i+1,j})\textrangle \\- \delta\omega_d\textlangle \tau_{i,j}((1-\tau_{i-1,j})\tau_{i+1,j}+(1-\tau_{i+1,j})\tau_{i-1,j})\textrangle\\-\delta^2\omega_d\textlangle \tau_{i-1,j}\tau_{i,j}\tau_{i+1,j}\textrangle
\mp \omega\textlangle \tau_{i,A}\tau_{i+1,A}(1-\tau_{i,B})\textrangle, \label{eq1}\\
\end{split}
\end{equation}
where $\textlangle\cdots\textrangle$ denotes the statistical average and last term on right-hand side takes a negative (positive) sign for lane-A (B).
At boundaries ($i=1, N$), the occupancy of particles evolve according to
\begin{eqnarray}
\frac{d\textlangle \tau_{1,j}\textrangle}{dt}=\alpha\textlangle (1-\tau_{1,j})\textrangle - \textlangle \tau_{1,j}(1-\tau_{2,j})\textrangle, \label{eq3}\\
\frac{d\textlangle \tau_{N,j}\textrangle}{dt}=\textlangle \tau_{N-1,j}(1-\tau_{N,j})\textrangle - \beta\textlangle \tau_{N,j}\textrangle. \label{eq4}
\end{eqnarray}
In the above system of equations correlations are neglecting by using mean-field approximation consist of the following approximation.
\begin{equation}
\textlangle \tau_{i,j}\tau_{i+1,j}\textrangle\approx\textlangle \tau_{i,j}\textrangle \textlangle \tau_{i+1,j}\textrangle. \label{eq5}
\end{equation}

For large system size $N\rightarrow \infty $ and small lattice spacing $\epsilon=L/N \rightarrow 0 $ with finite $N\epsilon$, one can derive the continuum
limit by coarse-graining a discrete lattice with quasi continuous variable $x=iL/N$ and rescaling the time as $t{'}=tL/N$. For simplicity, the length of the lattices is fixed as $L=1$ which restricts the variable $x$ in the range $[0,1]$. The time rescaling is useful to understand the engagement between particle conserving and non-conserving dynamics as the system attains stationary state locally due to conservative dynamics only and the non-conserving processes occur at a comparatively lower rate than particle conserving processes. To study the competition between boundary and bulk dynamics, we define the reduced the attachment, detachment and lane-changing rates inversely proportional to the system size~\cite{parmeggiani2003phase,parmeggiani2004totally}.
\begin{equation}
\Omega_a=\omega_a N, \Omega_d=\omega_d N, \Omega=\omega N.\label{eq6}
\end{equation}
Note that the parameters $\Omega_a$, $\Omega_d$ and $\Omega$ remain
fixed as $N\rightarrow \infty $.

In the continuum limit average density is replaced by a continuous variable by substituting $\textlangle \tau_{i,j} \textrangle\equiv \rho_{i,j}\in[0,1]$ and retain the terms up to $O(N^{-2})$ in Taylor's series expansion (for large system i.e. $N>>1$) to obtain
\begin{equation}
\rho_{i,j\pm 1}=\rho_{i,j}\pm \frac{1}{N}\frac{\partial\rho_{i,j}}{\partial x}+ \frac{1}{2N^2}\frac{\partial^2\rho_{i,j}}{\partial x^2}+ O\bigg(\frac{1}{N^3}\bigg).\label{eq7}
\end{equation}
Without loss of generality, we drop the subscript $i$ as the system is free form any kind of spatial inhomogeneity. The continuum mean field equations describing the steady state average densities ($\rho_A$ and $\rho_B$) in the bulk is given as
\begin{equation}
\frac{\partial }{\partial t{'}}\begin{bmatrix} \rho_A\\ \rho_B \end{bmatrix}+ \frac{\partial}{\partial x}\begin{bmatrix} -\frac{\epsilon}{2}\frac{\partial \rho_A}{\partial x}+ \rho_A(1-\rho_A)\\ -\frac{\epsilon}{2}\frac{\partial \rho_B}{\partial x}+ \rho_B(1-\rho_B) \end{bmatrix}=S, \label{eq8}
\end{equation}
where\\
$S= \begin{bmatrix} \Omega_a (1-\rho_A)(1+\rho_A(\delta-1))^2-\Omega_d \rho_A(1+\rho_A(\gamma-1))^2\\-\Omega \rho^{2}_A (1-\rho_B) \\ \Omega_a (1-\rho_B)(1+\rho_B(\delta-1))^2-\Omega_d \rho_B(1+\rho_B(\gamma-1))^2\\+\Omega \rho^{2}_A (1-\rho_B) \end{bmatrix}$.

Here, the term $S$ originates from the combination of particle lane changing transitions and modified adsorption-desorption kinetics and is responsible for the loss of particle conservation in the bulk of each lane. The components of $J$ represent the
currents in the particle conservation situation in lane-$A$ and $B$, respectively. The important noteworthy aspect
arising due to the biased lane-changing rule is that the coupling term acts as a sink for lane-A and a source for lane-B~\cite{gupta2014asymmetric}. Additionally, the particle-hole symmetry does not hold for the model with mutual interaction.
\section{Steady-state solution}
\label{sec:2}
In this section, we analyze the coupled system obtained form mean-field description under steady-state and discuss the effects of important parameters on density profiles and phase diagrams. The system \eqref{eq8} in the steady-state reduces to
\begin{eqnarray}
\begin{split}
\frac{\epsilon}{2}\frac{d^2 \rho_A}{d x^2}+(2\rho_A-1)\frac{d\rho_A}{dx}+\Omega_a (1-\rho_A)(1+\rho_A(\delta-1))^2\\-\Omega_d \rho_A(1+\rho_A(\gamma-1))^2-\Omega \rho^{2}_A (1-\rho_B)=0, \\\label{eq9}
\frac{\epsilon}{2}\frac{d^2 \rho_B}{dx^2}+(2\rho_B-1)\frac{d\rho_B}{dx}+\Omega_a (1-\rho_B)(1+\rho_B(\delta-1))^2\\
-\Omega_d \rho_B(1+\rho_B(\gamma-1))^2+\Omega \rho^{2}_A (1-\rho_B)=0. \label{eq10}
\end{split}
\end{eqnarray}

The boundary conditions for coupled nonlinear system \eqref{eq9} are $\rho_A(0)=\rho_B(0)=\alpha$, $\rho_A(1)=\rho_B(1)=1-\beta$. In the continuum limit $\epsilon \rightarrow 0$, the coupled system becomes overdetermined as the second order nonlinear differential equations reduce to first order differential equations keeping all the four boundary conditions intact. Here, the leading order terms in the above system is analogous to the dissipative term of viscous Burgers' equation in stationary state and play a similar role as performed by the vanishing viscosity term in the solution of Burgers' equation.

%However, retaining the second order terms ensures to generate a smooth solution fitting all the four boundary conditions. \textcolor[rgb]{1.00,0.00,0.00}{\emph{Due to the coupling term $\Omega \rho^{2}_A (1-\rho_B)$, the analytic solution for the average densities in both the lanes can not be obtained explicitly.}}

To understand the steady-state behavior of the system \eqref{eq8}, we employ
boundary layer analysis on the continuum mean-field equations. Recently, this approach has been quite successful in solving the hydrodynamic equation in the thermodynamic limit and explaining the complete rich phase diagrams of single-channel TASEP with LK~\cite{mukherji2006bulk} as well as two-channel TASEP with LK in
 fully and partially asymmetric coupling conditions~\cite{gupta2014asymmetric,dhiman2014effect}, respectively. The density profiles are constructed by computing the bulk part of the solution (outer solution) and boundary layer solution (inner solution) separately and then to match the both solutions suitably to get the global solution. The outer solution is the major part of the density profile and can be obtained in the thermodynamic limit ($L>>1$) due to the negligible contribution of the regularizing terms. We use a suitable numerical scheme to get approximate outer solution of the continuum
mean-field equations. The following numerical scheme for $j^{th}$ lane is used to
find the outer solution of continuum mean-field equations in both the lanes. \begin{eqnarray}
\begin{split}
\rho_{i,j}^{n+1} = & \rho_{i,j}^n+ \frac{\epsilon}{2}\frac{\Delta t{'}}{\Delta x^2}\big({\rho_{i+1,j}^n-2\rho_{i,j}^n+\rho_{i-1,j}^n}\big) \\&+
\frac{\Delta t{'}}{2\Delta x}\big[(2\rho_{i,j}^n-1)\big(\rho_{i+1,j}^n-\rho_{i-1,j}^n\big)\big]\\&+
\Delta t{'}\big[\Omega_a(1-\rho_{i,j}^n) (1+\rho_{i,j}(\delta-1))^2\\&-\Omega_d \rho_{i,j}^n (1+\rho_{i,j}(\gamma-1))^2\mp\Omega (\rho_{i,A}^n)^2(1-\rho_{i,B}^n)\big] \\&+ O(\Delta t{'}, \Delta x^2).
\end{split}
\end{eqnarray}
Here, the last term takes negative (positive) sign for lane-A (B), respectively. The stationary state density profiles have been obtained by capturing the solution
of above system in the limit $n >> 1$, which ensures the occurrence of steady state. This solution is also refereed as bulk solution. The solution describing the boundary layers or shocks can be obtained by ignoring the non-conservative terms in the steady-state system. This can be achieved by using a transformation $\widetilde{x}=\frac{x-x_d}{\epsilon}$, in the $\epsilon$-neighborhood of another position variable $x_d$ representing the position of boundary layer in the continuum mean field equations. This leads to the elimination of the source and sink terms in the hydrodynamic equations in regions of width of $O(\epsilon)$ and is well justified because particle non-conserving dynamics are irrelevant in regions of width of $O(\epsilon)$.  For more details of this approach we refer to \cite{mukherji2006bulk,gupta2014asymmetric,dhiman2014effect}.

%\emph{The omission of second-order system
% makes the coupled system over determined, due to which the outer solution is unable to
%meet the boundary conditions at both the boundaries, simultaneously. This generates the
%notion of left outer and right outer solutions. The solution satisfying left (right)
%boundary condition is known as left (right) outer solution. Since, density profile has
%to satisfy the boundary condition at other end also, the global solution can not be
%given by outer solution alone. So, to satisfy the boundary conditions at both the ends,
%a crossover narrow regime from left to right solution is formed which gives rise to
%either a boundary layer or a shock in the density profile. This solution is known as
%inner solution and is found by ignoring the non-conservative terms in the steady-state system.}

\section{Phase diagrams and density profiles}
\label{sec:3}
We restrict our investigation for a special choice $\Omega_a=\Omega_d$ to simplify the analysis and construct the phase diagrams. Importantly, the biased coupling of lanes makes the steady-state dynamics nontrivial in contrast to a symmetrically coupled two-lane TASEP with LK which is similar to two independent single-lane TASEP with LK ~\cite{gupta2014asymmetric}. In order to investigate the effect of mutually interactive LK dynamics on the steady-state properties, the whole discussion is divided into two parts based on the symmetric or antisymmetric LK rates. In the symmetric case, attachment and detachment rates are modified in a similar fashion while in the antisymmetric case, if attachment rate is enhanced than the detachment rate is reduced by the same amount and vice-versa. The results obtained from continuum mean-field equations are validated with Monte-Carlo simulations utilizing random sequential update. In the simulations, a lattice of size $L=1000$ is used to minimize boundary effects and the simulations are carried out for $10^{10}$ time steps. To ensure the occurrence of steady-state, the first $5 \%$ steps are discarded and the densities in both the lanes are computed by taking time averages over an interval of $10 L$. The phase boundaries are calculated within an estimated error of less than $1\%$.
%The symmetry in coupling rates leads to the cancelation of lane-changing source terms with sink terms in the mean-field hydrodynamic equations, which gives two uncoupled ordinary differential equations representing two independent TASEPs with LK. Hence, the topology of phase diagram of single-lane TASEP with LK model remains preserved. This is totally in contrast to the fully asymmetric coupling conditions~\cite{gupta2013asymmetric}, where we have existence of a novel phase diagram, considerably different from that of a single-lane TASEP with LK~\cite{mukherji2006bulk, parmeggiani2004totally}.
\section{Case 1: Mutual interaction with symmetric LK rates}
In this section, we construct the density profiles and derive phase diagram for $\Omega=1$ and $\Omega_d=0.2$ under the mutually interactive symmetric LK dynamics. To investigate the effect of mutual interactions with symmetric modified LK, we choose $\gamma=\delta=1+\theta$ i.e. both the attachment and detachment rates are symmetrically modified with a multiple of  $1+\theta$ for each occupied neighboring sites with $\theta \geq -1$. Here, $\theta$ represents the strength of the mutual interaction under symmetric LK rates. It is to be noted that the positive (negative) values of $\theta$ increases (decreases) the LK dynamics provided the neighboring sites are occupied.\\
%%%%%%%%%%%%%%%%%%%%%%%%%%%%%%%%%%%%%%%%%%%%%%%%%%%%%%%%%%%%%%%%%
\begin{figure}
\includegraphics[trim=20 00 30 00,width=4.25cm,height=4.00cm]{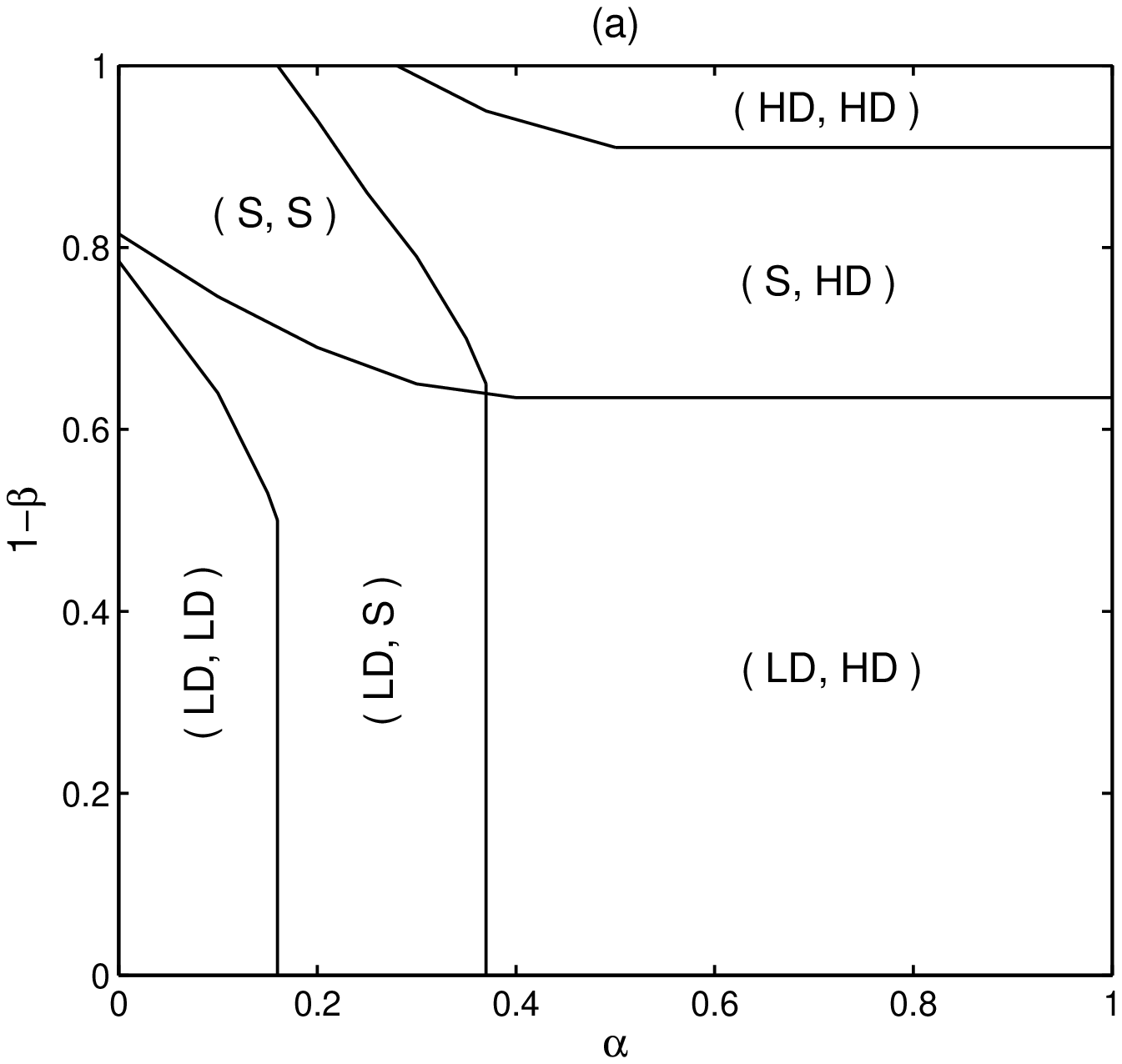}
\includegraphics[trim=20 00 30 00,width=4.25cm,height=4.00cm]{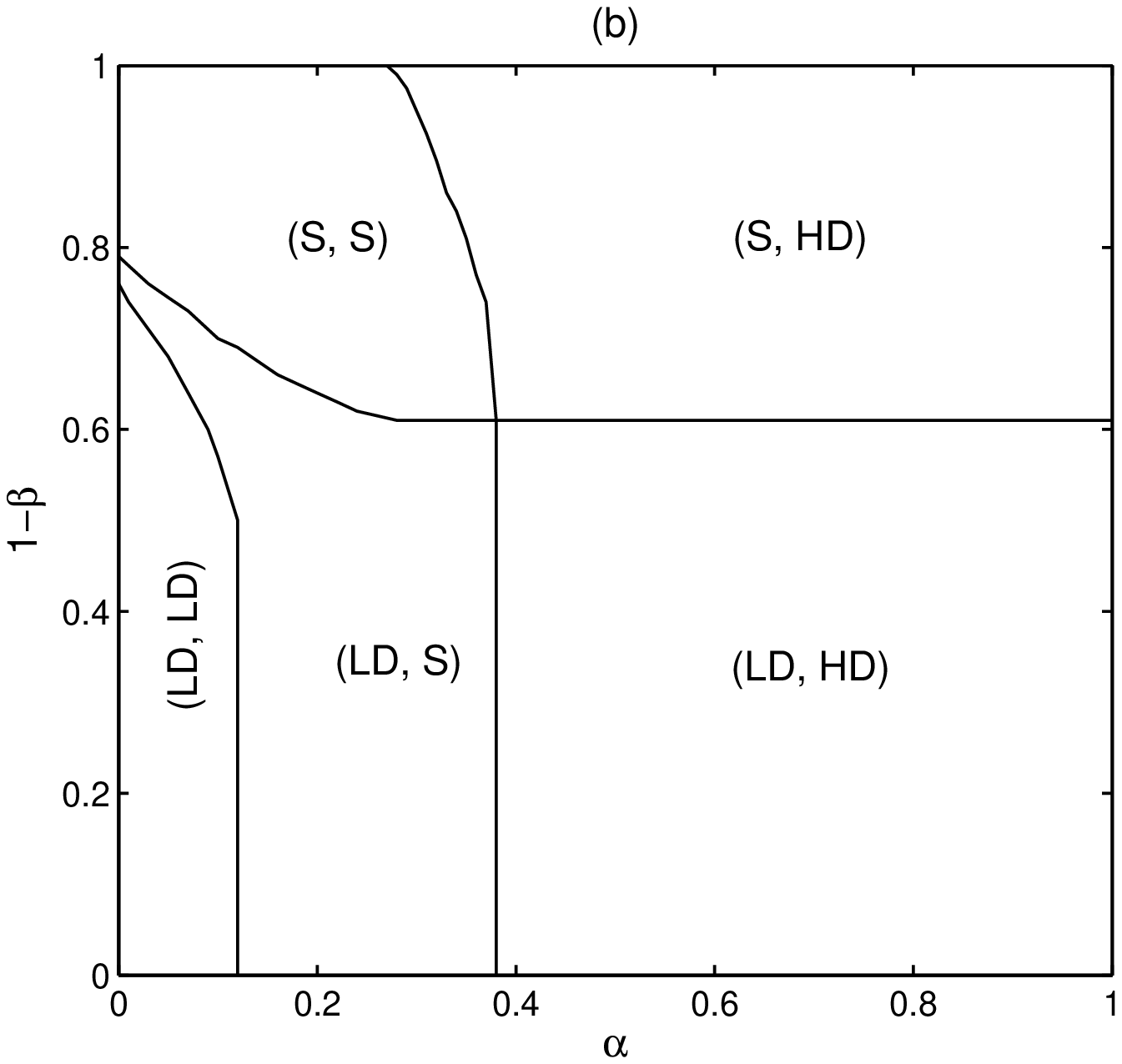}\\
\includegraphics[trim=20 00 30 00,width=4.25cm,height=4.00cm]{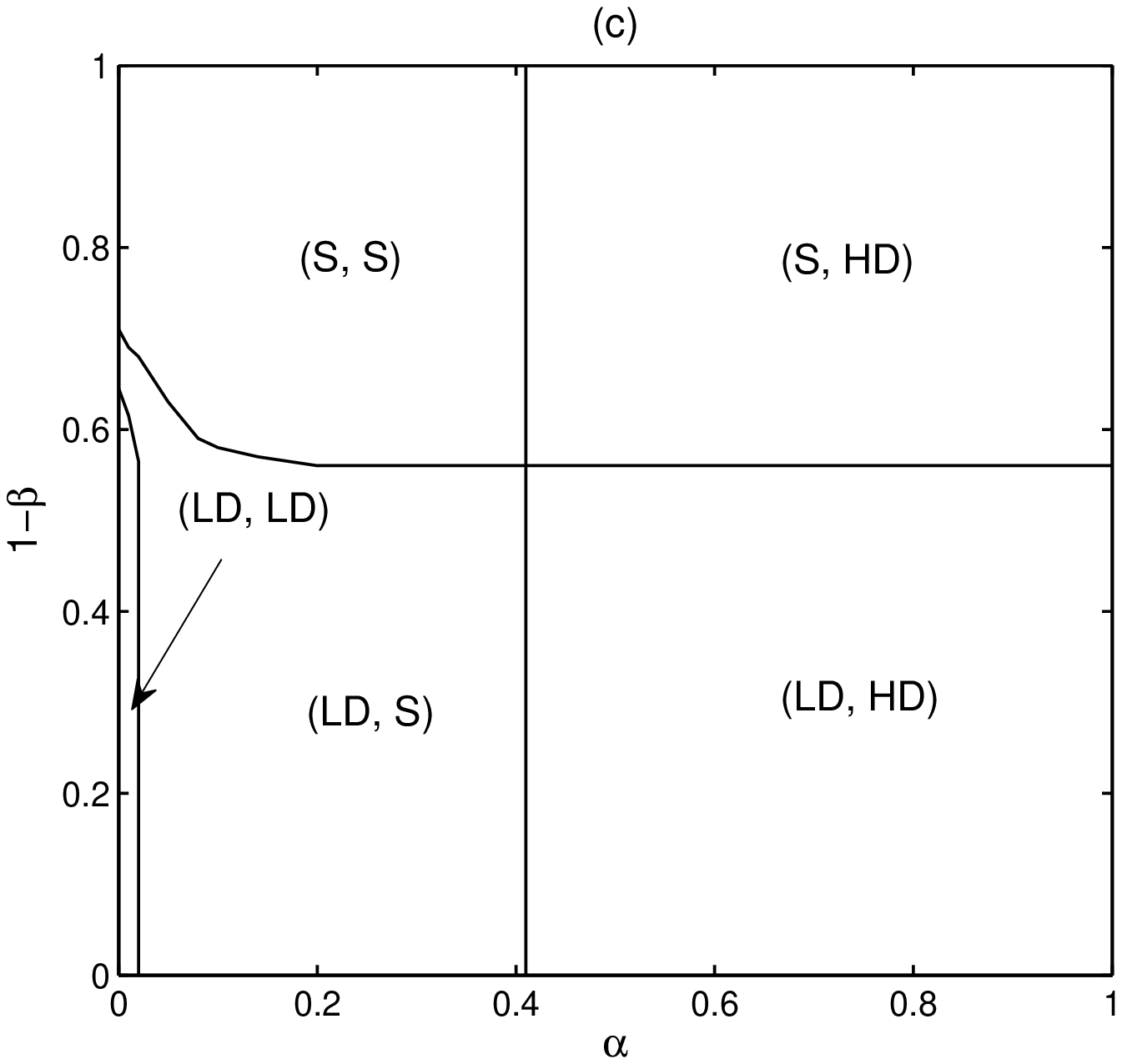}
\includegraphics[trim=20 00 30 00,width=4.25cm,height=4.00cm]{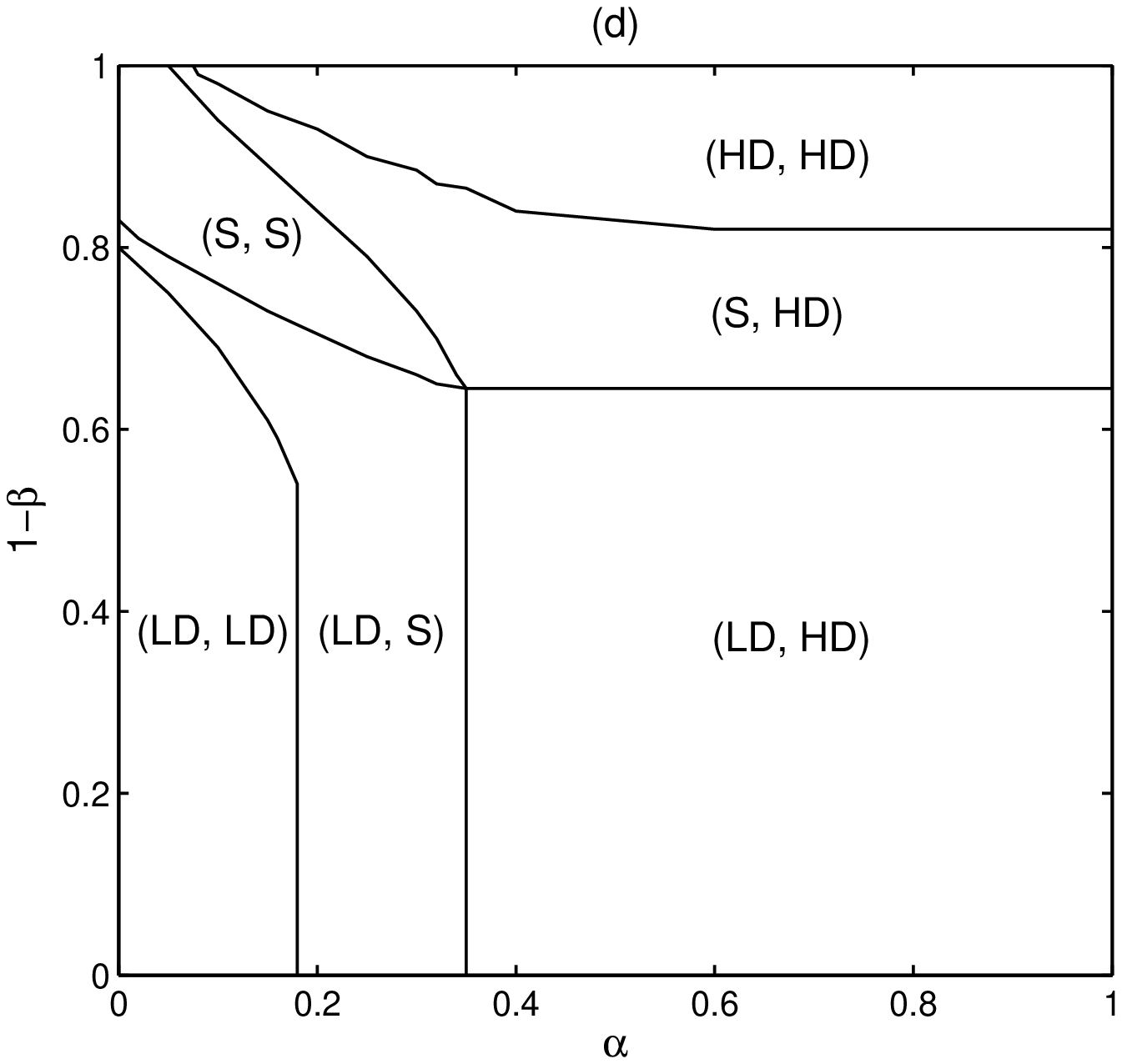}\\
\includegraphics[trim=20 00 30 00,width=4.25cm,height=4.00cm]{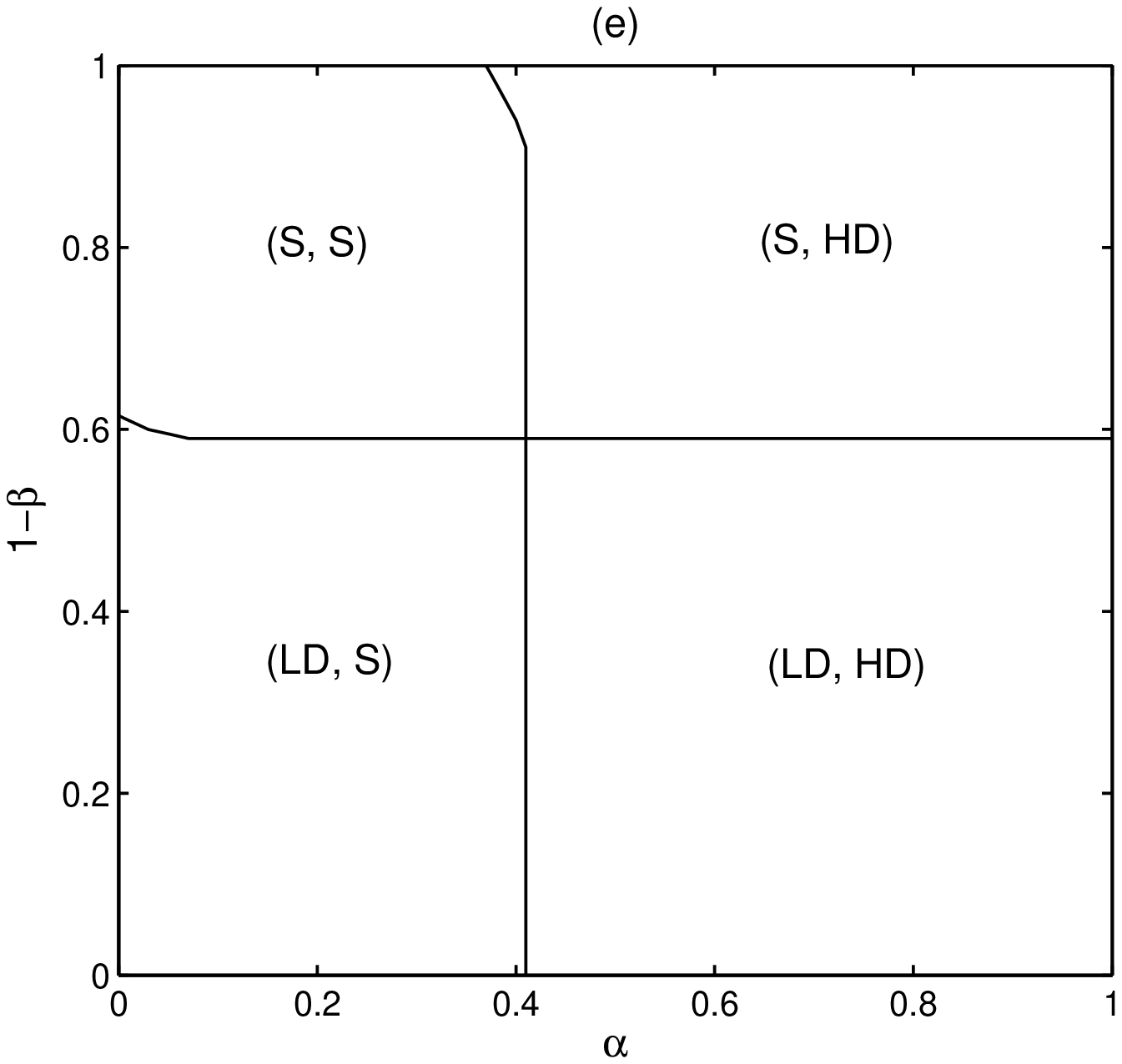}
 \caption{\label{fig:2}(Color online) Phase diagrams for $\Omega=1$. (a) $\theta=0$ i.e. no mutual interaction, (b) $\theta=0.5$, (c) $\theta=2.0$, (d) $\theta=-0.5$, (e) $\theta=0$. In (a)-(d): $\Omega_d=\Omega_a=0.2$ and in (e): $\Omega_d=\Omega_a=0.5$.  Here, LD, HD and S denotes Low density, High density and Shock phase, respectively.
}
\label{fig:2}
\end{figure}
%%%%%%%%%%%%%%%%%%%%%%%%%%%%%%%%%%%%%%%%%%%%%%%%%%%%%%%%%%%%%%%%%%
For $\theta = 0$, the present case will convert in to a simple asymmetrically coupled two-lane TASEP with the LK system without mutual interaction whose steady-state behavior has been recently investigated \cite{gupta2014asymmetric} and found to be quite complex as compared to the symmetric coupling. It has been reported that a small asymmetry in lane changing rates can produce significant changes in the phase diagram without mutually interactive LK \cite{pronina2006asymmetric}. Figure 2(a) shows the phase diagram in terms of bulk transitions with $\theta = 0$, which has been adapted from ref.~\cite{gupta2014asymmetric} and is provided here for the sake of comparison. Since the surface transitions are responsible only for the change in the slope of the boundary layer, we for simplicity, focus on the simpler representation of the phase diagram showing only the bulk transitions in the phase plane. Moreover, this simpler structure depicting the qualitative nature of phases is helpful in analyzing the effect of mutual interactive LK on the phase diagram. Broadly classifying, there exists six steady-state distinct phases viz. (LD,LD), (LD,S), (LD,HD), (S,S), (S,HD), and (HD,HD) as shown in fig.~\ref{fig:2}(a). Due to the asymmetric coupling between the lanes, lane A acts as a homogeneous bulk reservoir for lane B. This creates an imbalance between adsorption and desorption rates in both the lane leading to a higher effective detachment (attachment) rate for lane A (B). As a result, major region of the phase diagram is filled with LD (HD) phase for lane A(B). This combined effect can be seen in the phase diagram of individual lanes A and B which are similar to the one channel TASEP with LK for $\Omega_d>\Omega_a$ and $\Omega_a>\Omega_d$, respectively.

To investigate the effect of mutual interaction with the background, the phase diagrams for nonzero $\theta$ are shown in figs.~\ref{fig:2}(b)-(d). It is clear from the figures that the (HD, HD) phase disappears quickly from the phase diagram while the (LD, LD) phase shrinks slowly on increasing $\theta$. This is due to the fact that for higher $\theta$ there will be more detachments (attachments) in HD (LD) phase due to the higher probability of occupied neighboring sites. For $\theta=0.5$, the number of steady-state phases reduces to five, consisting of all the phases except (HD, HD) as compared to the one with $\theta=0$. For $\theta = 2$, phase (LD, LD) becomes confined to a very small region near to the boundary $\alpha = 0$ (fig.~\ref{fig:2}(c)). The reverse phenomenon happens when we start decreasing $\theta$ from zero. The phases (LD, LD) and (HD, HD) expand while the other phases shrinks. Note that there does not emerge any new phase for negative values of $\theta$.

Under the symmetric LK rates, the effect of $\theta$ on the phase diagram is found analogous to the one with $\Omega_a$. To compare the effect of $\theta$ with $\Omega_a$, the phase diagram for higher value of $\Omega_a = 0.5$ for $\theta=0$ is drawn in fig.~\ref{fig:2}(e). The topology of the phase diagram for increasing $\theta$ is similar to the one corresponding to increasing $\Omega_a$ with the only exception that (LD, LD) phase still present in a narrow region for higher $\theta$. The main difference between varying $\theta$ and $\Omega_a$ is that the impact of $\theta$ is much stronger in the (HD, HD) phase as compared to $\Omega_a$.

%%%%%%%%%%%%%%%%%%%%%%%%%%%%%%%%%%%%%%%%%%%%%%%%%%%%%%%%%
\begin{figure}
\includegraphics[clip,width=6.25cm,height=5.5cm]{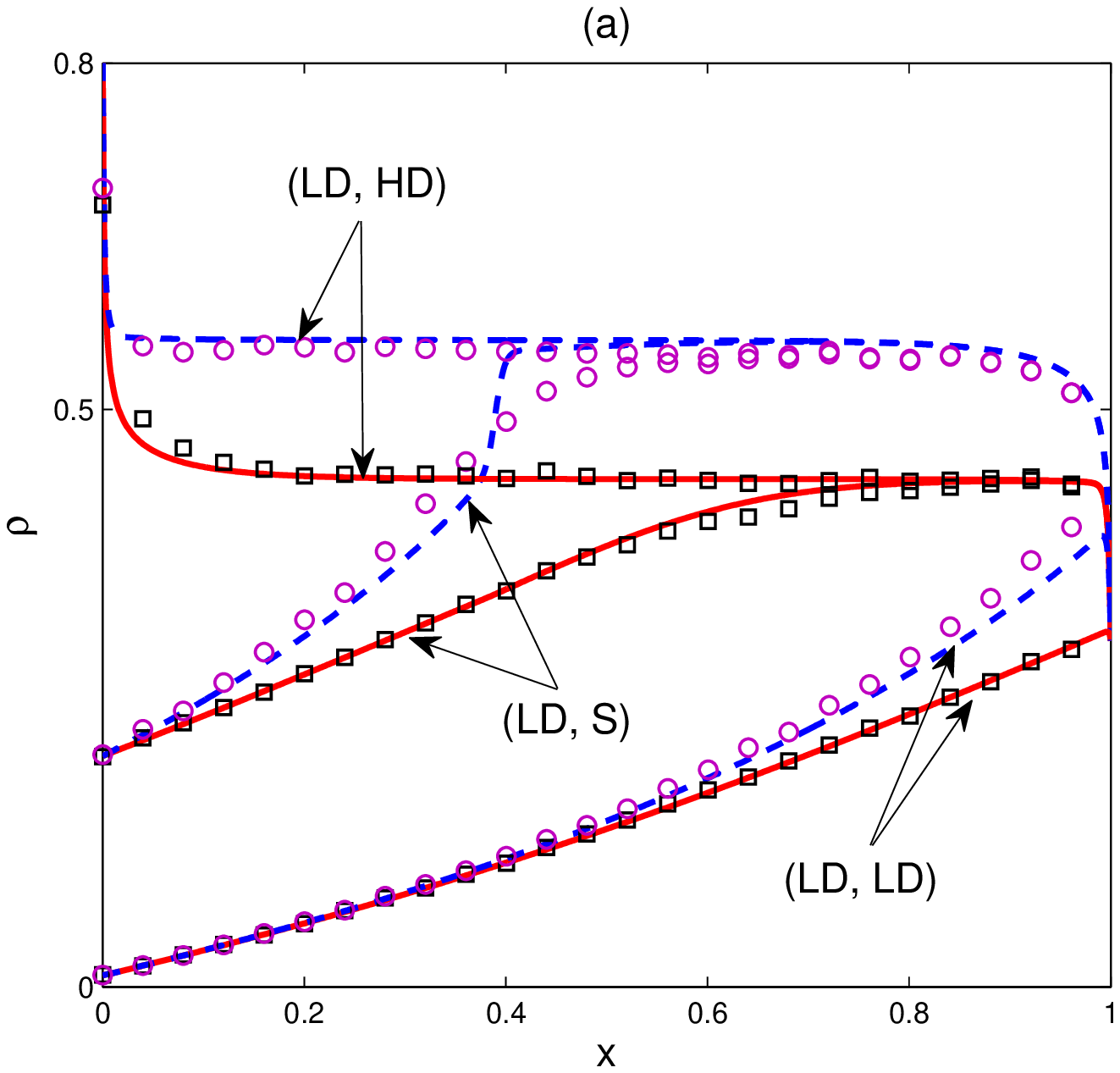}
\includegraphics[clip,width=6.25cm,height=5.5cm]{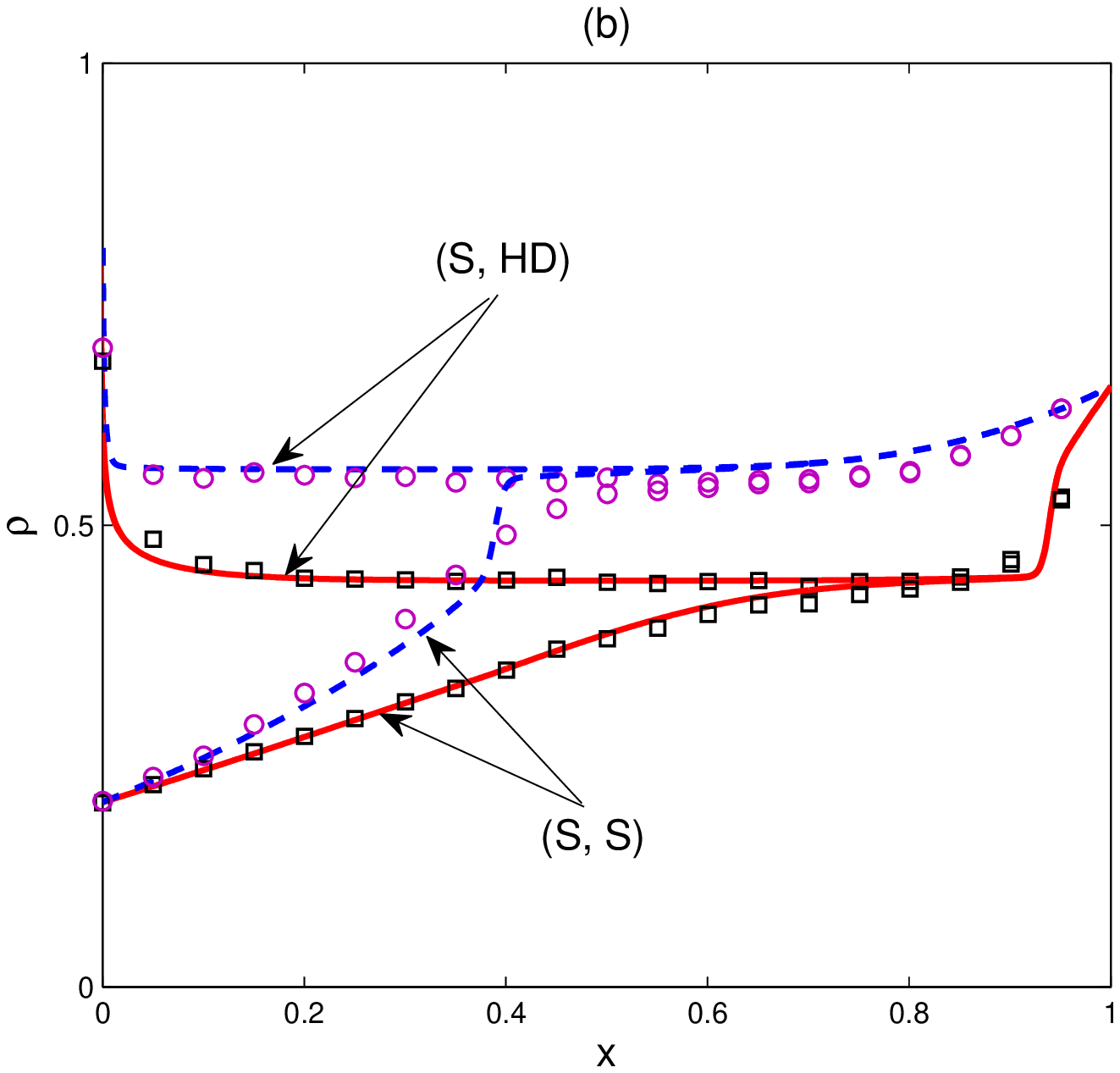}

 \caption{(Color online) Density profiles in symmetric case with $\theta=2.0$ and $\Omega_d=\Omega_a=0.2$ (a) (LD, LD) for $\alpha=0.03$ and $\gamma=0.3$; (LD, S) for $\alpha=0.2$ and $\gamma=0.3$; and (LD, HD) for $\alpha=0.8$ and $\gamma=0.3$, (b) (S, S) for $\alpha=0.2$ and $\gamma=0.65$; and (S, HD) for $\alpha=0.8$ and $\gamma=0.65$. The solid (dashed) lines in red (blue) color are the continuum mean-field density profiles of lane A (B). The curves marked with squares (circles) are the result of Monte Carlo simulations for lane A (B).}\label{fig:3}
\end{figure}

\begin{figure}
\includegraphics[width=6.25cm,height=5.5cm]{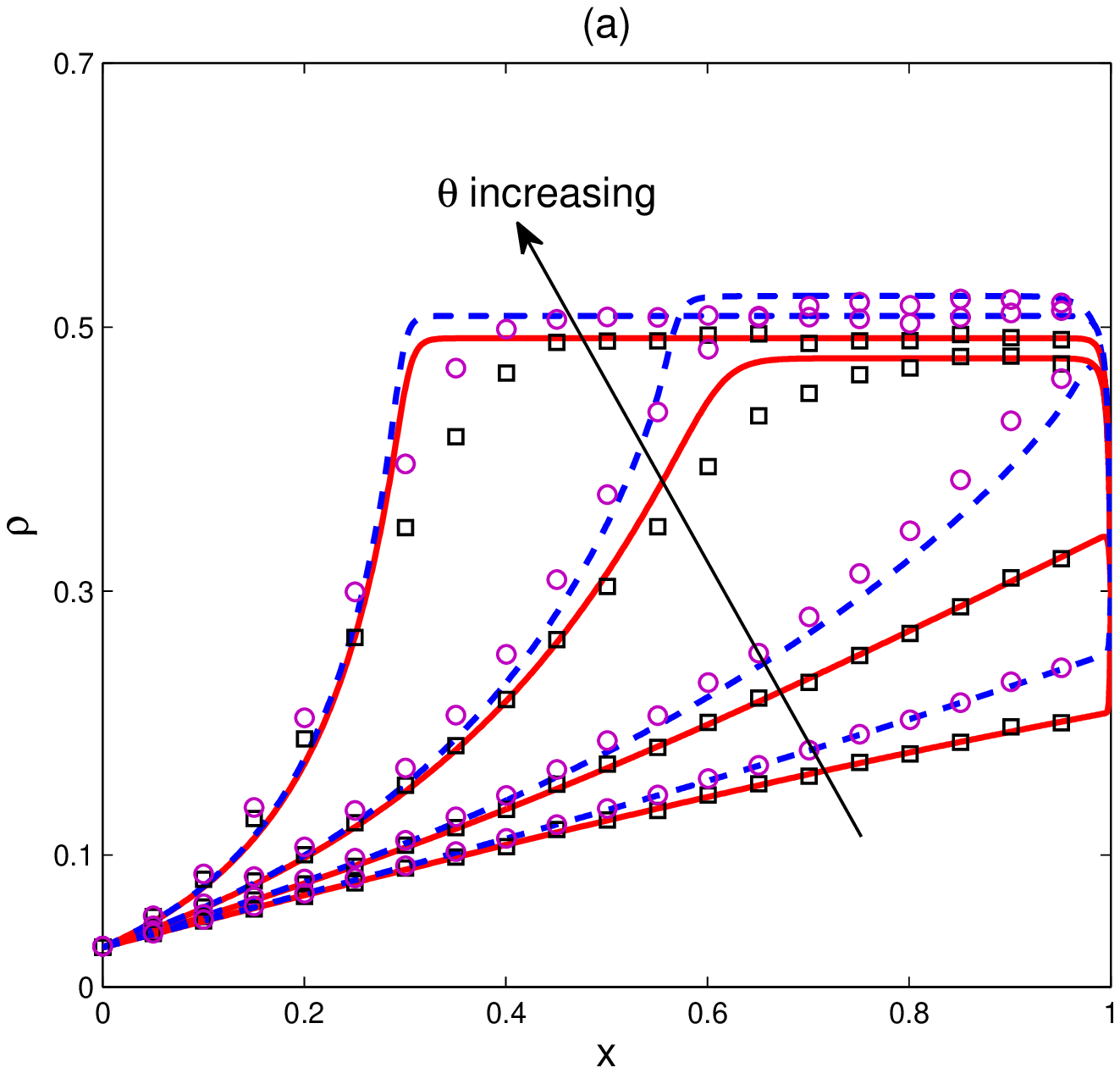}
\includegraphics[width=6.25cm,height=5.5cm]{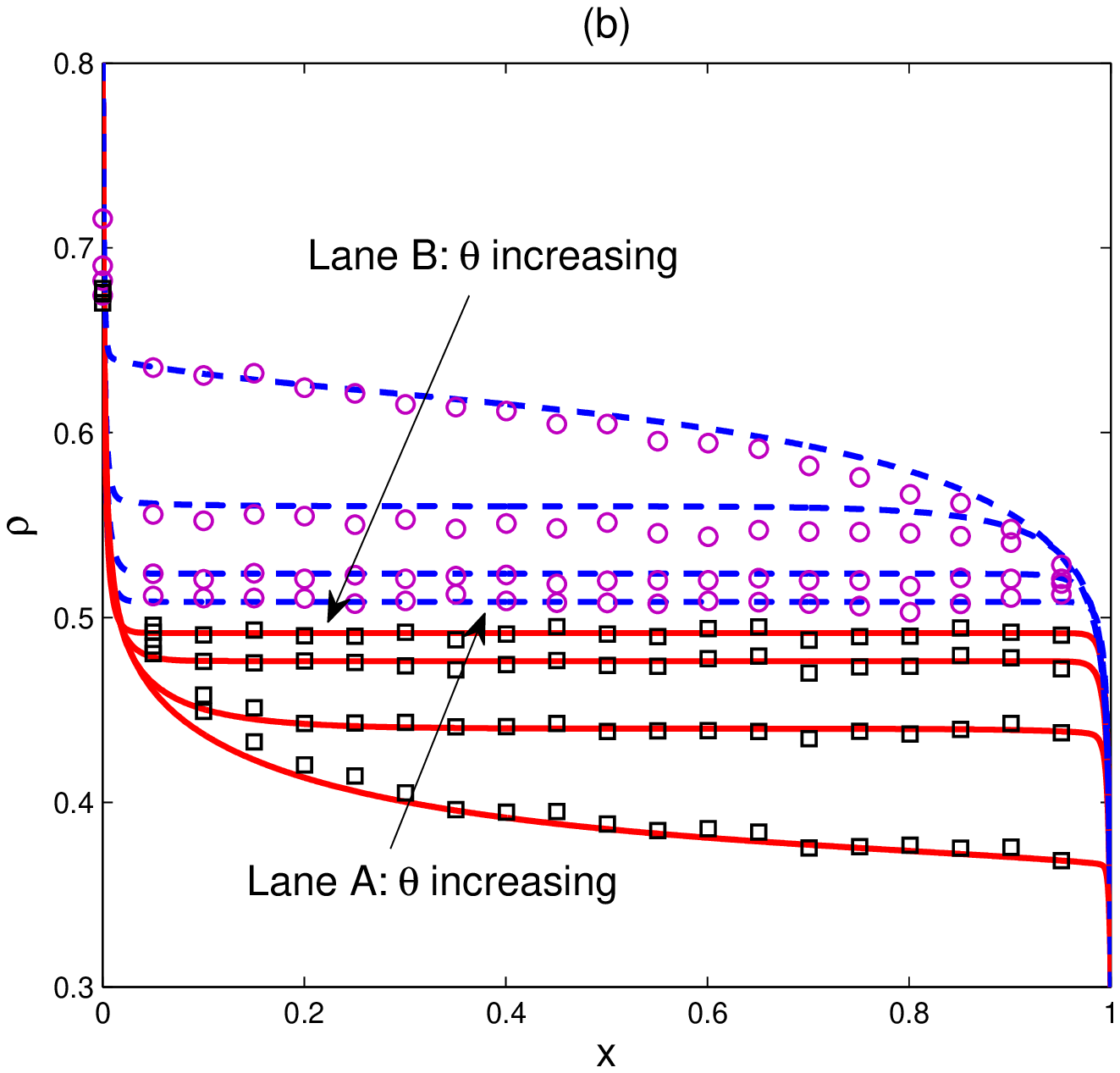}

 \caption{(Color online) Effect of $\theta$ on density profiles in both the lanes for $\theta = 0, 2, 5, 10$ (a) $\alpha=0.03$ and $\gamma=0.3$, (b) $\alpha=0.8$ and $\gamma=0.3$. The continuum mean-field and Monte Carlo simulation results are represented, respectively, by solid (dashed) lines in red (blue) color and squares (circles) for lane A (B).}\label{fig:4}
\end{figure}

Fig.~\ref{fig:3} shows the density profiles of all five different phases for $\theta = 2.0$ from continuum mean-field equations and their validation using Monte Carlo simulations. In general, there is a good agreement between the simulation and theoretical results except near the boundary layer due to the finite size effect. It is to be noted that $\rho_A < \rho_B$ because of the fully asymmetric coupling as particles of lane-B cannot move to lane-A. The effect of $\theta$ on (LD, LD) phase is shown in fig.~\ref{fig:4}(a). From the variation of density profile with respect to $\theta$, it is clear that the phase (LD, LD) disappear with an increase in $\theta$ and converts in to mixed type of phase in lane B for larger values of $\theta$. The mixed type of phase corresponds to the two-phase coexistence of LD with HD. It is to be noted that the LD-HD phase is not characterized by a discontinuity or a shock localized in the bulk of lane B. This is due to the fact that the LD and HD profiles are not straight lines for $\theta>0$ which is in contract to the case of $\theta=0$. Moreover, the density profile in both the lanes, regardless of the phase, approach to Langmuir isotherm as $\theta$ increases. In the limit of $\theta\rightarrow\infty$, the Langmuir Kinetics dominates the system dynamics completely and the density in both the lanes becomes Langmuir density i.e. $\rho_A=\rho_B=1/2$. Further, the effect of $\theta$ on the (LD, HD) phase is examined in fig.~\ref{fig:4}(b). The existence of (LD, HD) phase for larger $\theta$ is shown. In contrast to the case of $\theta=0$, density becomes uniform throughout the lattice in both the lane for higher values of $\theta$ and approaches to Langmuir density.

\begin{figure}
\includegraphics[width=6.25cm,height=5.5cm]{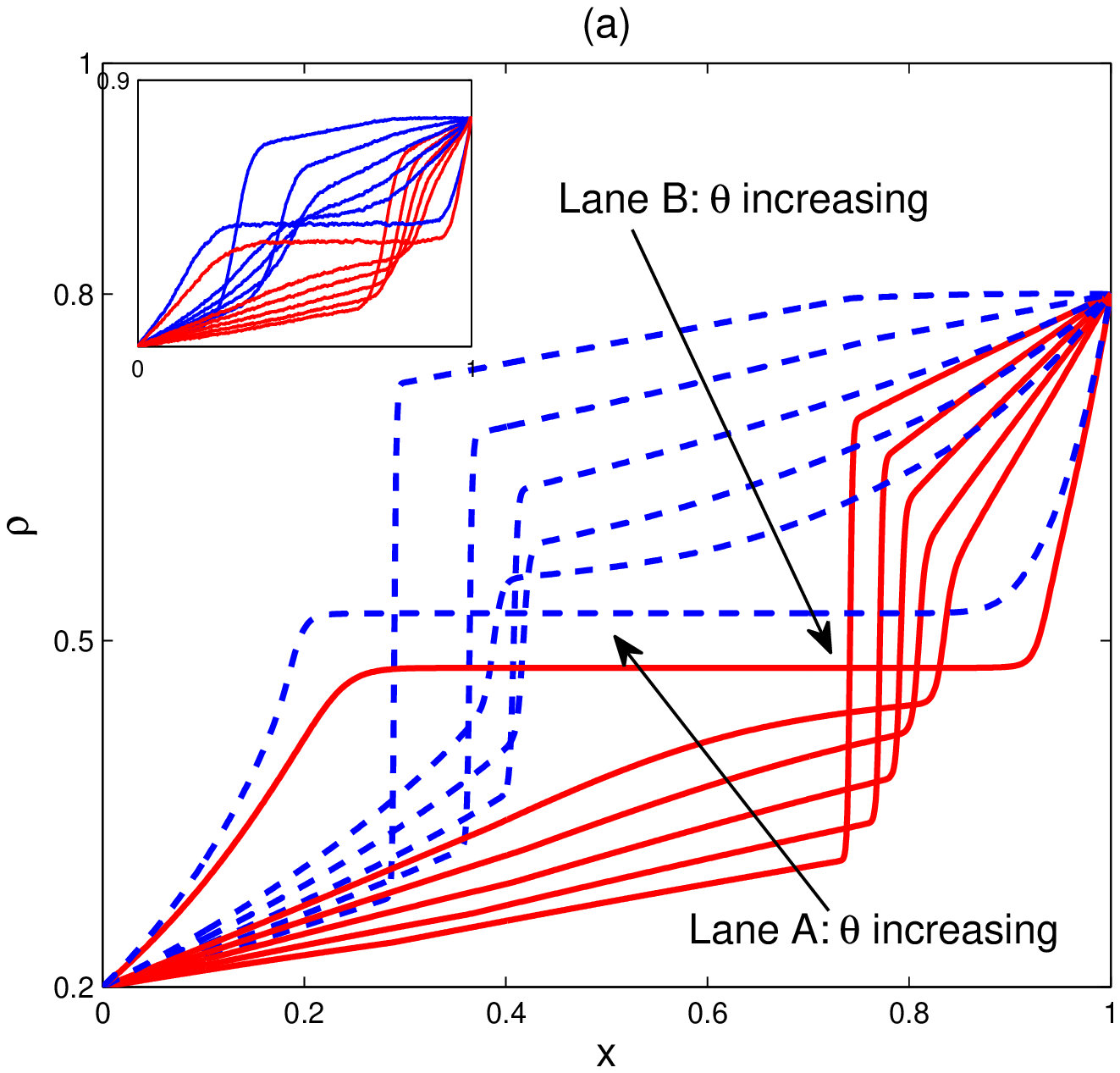}
\includegraphics[width=6.25cm,height=5.5cm]{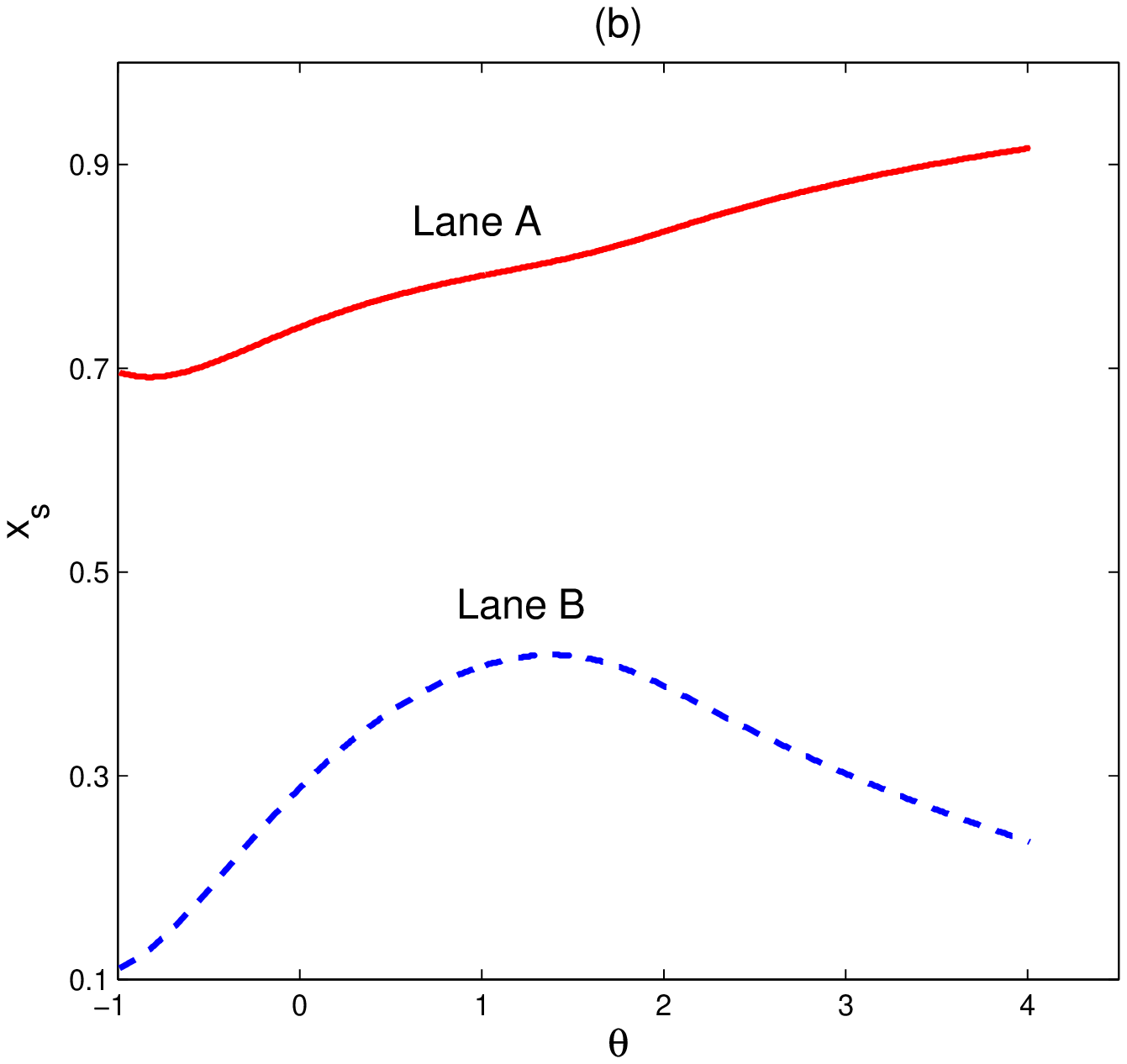}
 \caption{(Color online) Effect of $\theta$ on shocks in (S, S) phase for $\alpha=0.2$ and $\gamma=0.8$ (a) $\theta = 0, 0.5, 1.0, 1.5, 2.0, 5.0$. Inset shows verification via Monte Carlo simulation results for $N=1000$, (b) Location of shock $(x_s)$ vs. $\theta$ in both the lanes.} \label{fig:5}
\end{figure}

A shock is a discontinuity in bulk connected by a low and high density part. Across the boundary layer in the bulk, i.e. a shock, the current in the system remains constant \cite{mukherji2006bulk}. The effect of $\theta$ on the shock dynamics in the (S, S) phase is explored in Fig.~\ref{fig:5}(a). The first thing one can notice is that $\theta$ has a negative effect on the shock hight. The indistinguishable feature of the shocks in our system is that shock in lane A continuously moves towards right boundary on increasing $\theta$ while the shock in lane B shows a different trend known as jumping effect \cite{wang2007effects}. As shown in fig.~\ref{fig:5}(b)  shock in lane-B moves to the right first till $\theta \leq 1.5$ and  changes it's direction for further increase in $\theta$ and starts moving towards left boundary for $\theta > 1.5$. Moreover, for comparatively larger value of $\theta$ shock in both the lanes disappears and bulk density of the whole lattice approaches to Langmuir density.
%%%%%%%%%%%%%%%%%%%%%%%%%%%%%%%%%%%%%%%%%%%%%%%%%%%%%%%%%

%%%%%%%%%%%%%%%%%%%%%%%%%%%%%%%%%%%%%%%%%%%%%%%%%%%%%%%%%

%The two-lane TASEP with LK model has already been analyzed under symmetric ~\cite{wang2007effects} as well as fully asymmetric coupling conditions ~\cite{gupta2013asymmetric}. Now, we consider the important case of partially  asymmetric coupling conditions, which has been unexplored yet.
%It is reported in the literature ~\cite{pronina2004two, pronina2006asymmetric, Gupta20136314, shi2011strong} that the phase diagram of a two-lane TASEP without LK is significantly different in partially asymmetric and fully asymmetric coupling environments. This stimulates the need to answer two important questions:
%(a) Does there exist any difference in the phase diagrams of a two-lane TASEP with LK under partially asymmetric and
%fully asymmetric coupling environments? (b) If there exist any differences, are these of similar kind as
%observed in the corresponding system without LK or not?

\section{Case 2: Mutual interaction with antisymmetric LK rates}
So far, we have investigated the effect of modified LK rates when both attachment and detachment rates are enhanced or reduced simultaneously by equal amount while the hopping rate remains unchanged. Recently, in an in-vitro experiment, the presence of attractive interactions has been seen in kinesins-1, which remain longer attached to the microtubule in the presence of neighboring motors and leads to formation of motor clusters \cite{roos2008dynamic}. In this section, the proposed model is used to understand the coordination mechanism under the existence of attractive interactions. The effect of mutual interaction is analyzed in antisymmetric manner where the attachment (detachment) rate is enhanced (reduced) and vice versa.\\
 %%%%%%%%%%%%%%%%%%%%%%%%%%%%%%%%%%%%%%%%%%%%%%%%%%%%%%%%
\begin{figure}
\includegraphics[trim=20 00 30 00,width=4.25cm,height=4.00cm]{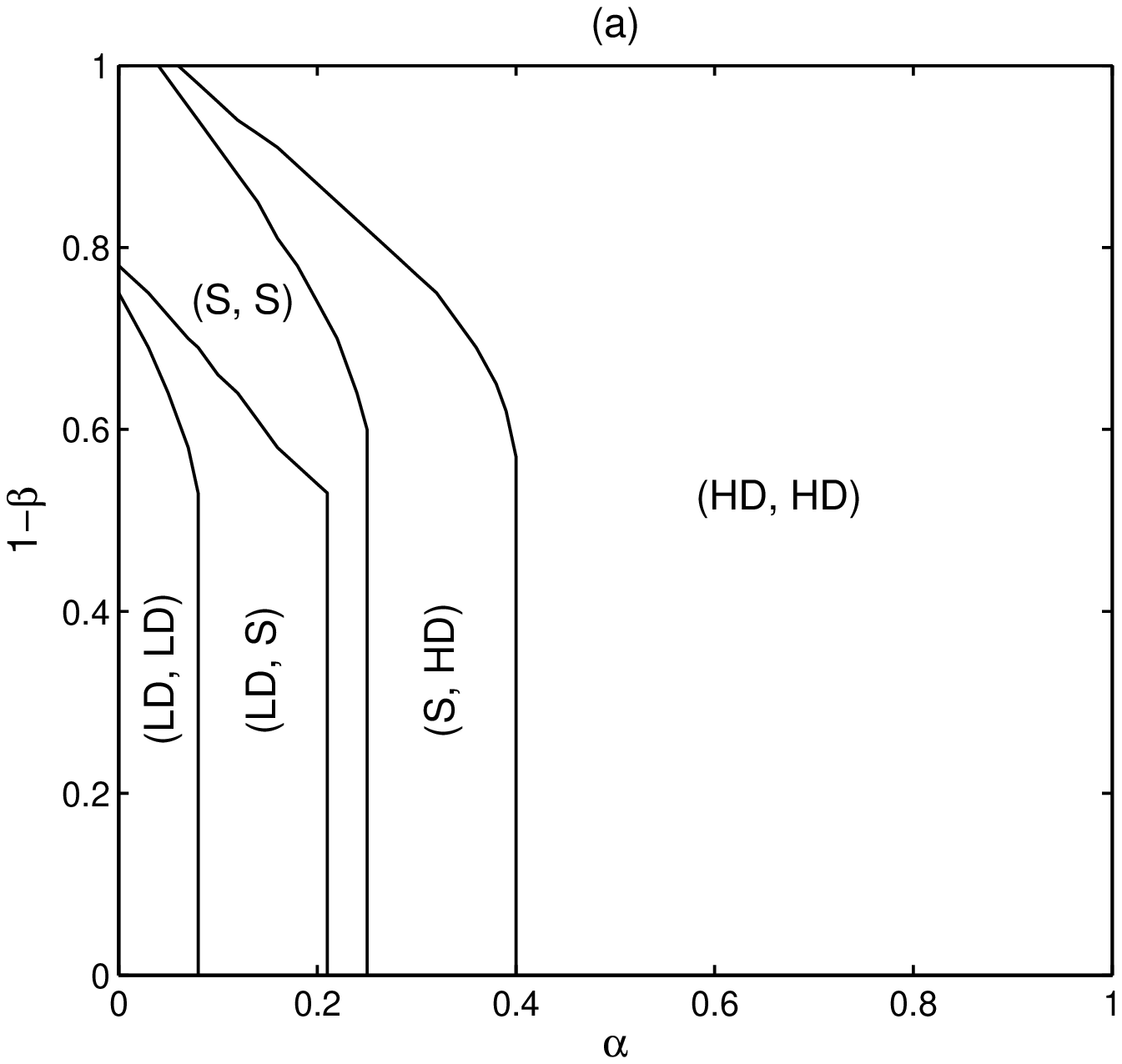}
\includegraphics[trim=20 00 30 00,width=4.25cm,height=4.00cm]{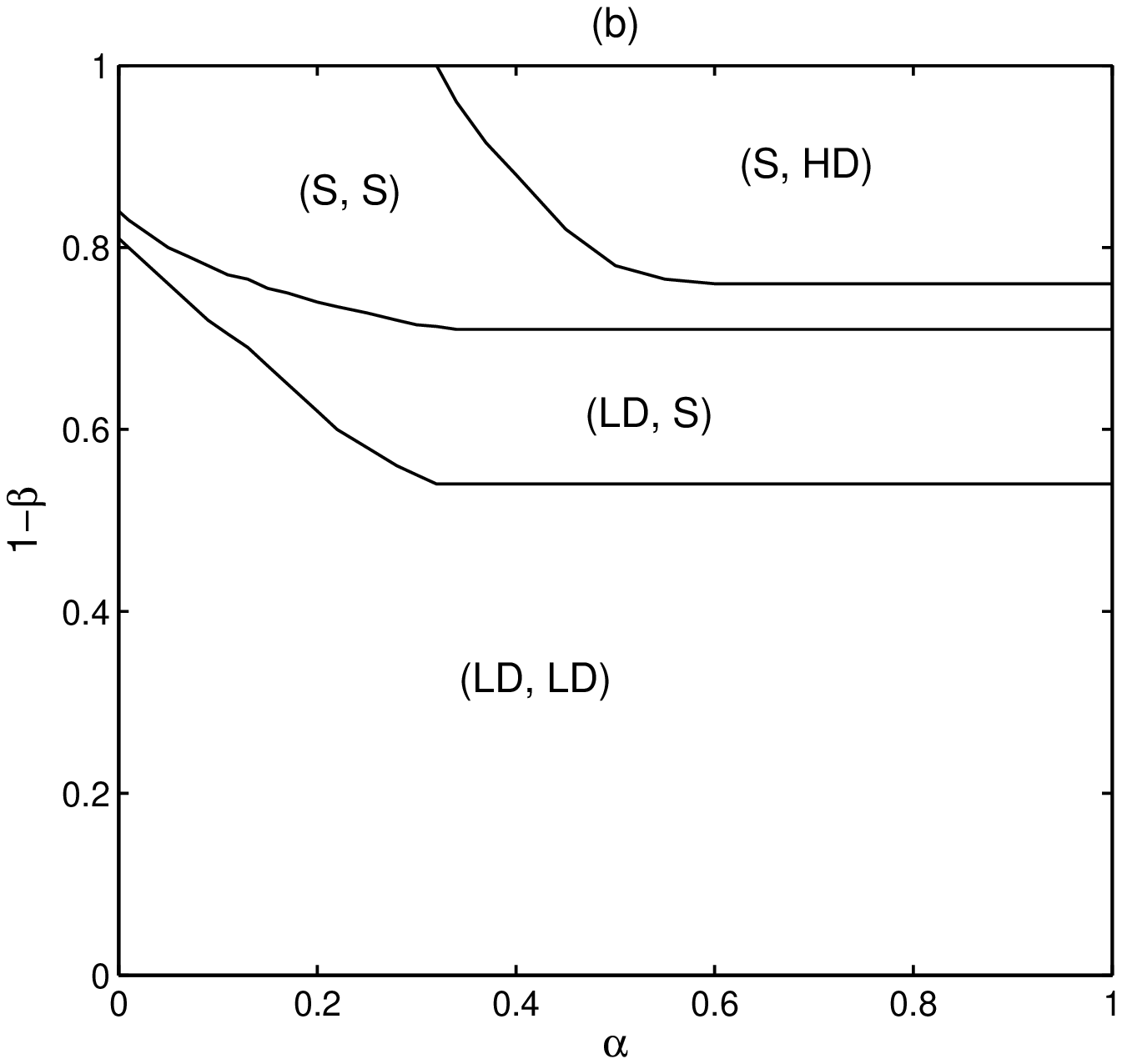}

 \caption{(Color online) Phase diagrams in antisymmetric case with $\Omega_d=\Omega_a=0.2$, and $\Omega=1$ for different values of $\phi$. (a) $\phi=0.5$, and (b) $\phi=-0.5$.}\label{fig:6}
\end{figure}
 To begin with we set $\gamma=1+\phi$ and $\delta=1-\phi$. Here, $\phi$ is a constant having the range [-1,1] as any value outside this interval leads to negative rate which is unrealistic. Positive values of $\phi$ represents the attractive interaction while negative values signify the repulsive interaction of the LK dynamics. The effect of $\phi$ on the density profiles are analyzed and phase diagrams are constructed using the methods discussed in previous section. Firstly, we investigate the role of attractive interactions on the phase diagram by increasing the value of $\phi$. For $\phi = 0$, the system has been well studied and have six distinct phases (see fig.~\ref{fig:2}(a))\cite{gupta2014asymmetric}. The increase in the value of $\phi$ firstly retains the topological structure of the phase diagram with six distinct phases while the phase boundaries are slightly shifted. Due to enhanced attachment rates via mutual interactions, the region containing (LD,LD) phase shrinks, while the region of (HD, HD) phase expands with increasing $\phi$. The significant changes in the topological structure of the phase diagram are observed for $\phi =0.5$ as shown in Fig.~\ref{fig:6}(a). Note that there does not emerge any new phase and the phase diagram consists of five phase having the same characteristics as for $\phi =0$. The (HD, HD) phase captures majority of the region in the phase diagram while the (LD, HD) phase completely disappears. In fig.~\ref{fig:7}(a), the density profiles for three different phases from continuum mean field along with the Monte Carlo simulation results are shown corresponding to the phase diagram in Fig.~\ref{fig:6}(a) with $\Omega =0.2$.

 The transition from (LD, HD) to (HD, HD) with increasing $\phi$ can be understood as follows (see fig.~\ref{fig:8}). Due to the increased number of particles attaching from the background in both the lanes, the density in both the lanes increase with $\phi$ which leads to abundance of particles in the system. As a result, very less number of particles shift from lane A to lane B for higher $\phi$. An noteworthy aspect of the density profile is the origination of a mixed type of profile for intermediate range of $\phi$, where some part of the bulk is in low density while the remaining part have high density. So, it is reasonable to conclude that LK dominates the lane changing dynamics in the presence of higher attractive interactions with the background.\\

 Now, we discuss the important and distinguishing features of the phase diagram for negative $\phi$. Though the repulsive interactions has not been studied in literature for motion of molecular motor, yet the effect of negative values of $\phi$ on the phase diagram has been explored for the sake of completeness as well as the results may be useful for explaining many other nonequlibrium phenomena. As contrast to the case of attractive interactions, (LD, LD) phase expands and covers major portion of the phase diagram for smaller values of $\phi$. This can be understood on the similar lines as discussed for the case of (HD, HD) phase for positive $\phi$. In fig.~\ref{fig:6}(b), the phase diagram for $\phi=-0.5$ are provided. There are only four different phases in the phase diagram which were also seen for the case of no mutual interactions i.e. $\phi=0$. It is to be noted that the topological structure of the phase diagram is quite different as compared to any other phase diagram discussed in this paper. Fig.~\ref{fig:7} shows the density profile of different phases computed form continuum mean field theory and compared with Monte Carlo simulation results. Although some discrepancies in the some part of bulk of Lane B is reported in Fig.~\ref{fig:7}, yet, there is a good agreement in general between the analytical and simulation results for antisymmetric LK dynamics. The discrepancies in Lane A occurs only near the boundary layer in the bulk i.e. shock and is due to finite size effect. On the other hand, whenever there is a shock in lane A, the discrepancies occur in the left part of Lane B form the position of the shock in Lane A. Moreover, the simulation results overshoot (undershoot) the analytical results for positive (negative) $\phi$ in the portion of discrepancies in Lane B. There are two reasons for such discrepancies, the finite lattice size and the mean field approximation. Moreover, the phase transition from (LD, HD) to (LD, LD) is shown in Fig.~\ref{fig:8}. Further, the effect of $\phi$ on the domain wall in (S, S) phase is analyzed in Fig.~\ref{fig:9}. In contrast to the symmetric case, shocks in both the lanes moves towards left boundary with increasing $\phi$.

 The above analysis reveals that the mutual interacting LK dynamics has an important effect on the steady-state behavior of the two-channel system. Due to the recent findings on clustering of motor proteins on microtubules, this effect has been modeled by modifying the LK dynamics based on the nearest neighbor interactions.

  %%%%%%%%%%%%%%%%%%%%%%%%%%%%%%%%%%%%%%%%%%%%%%%%%%%%%%%%

\begin{figure}
\includegraphics[width=6.25cm,height=5.5cm]{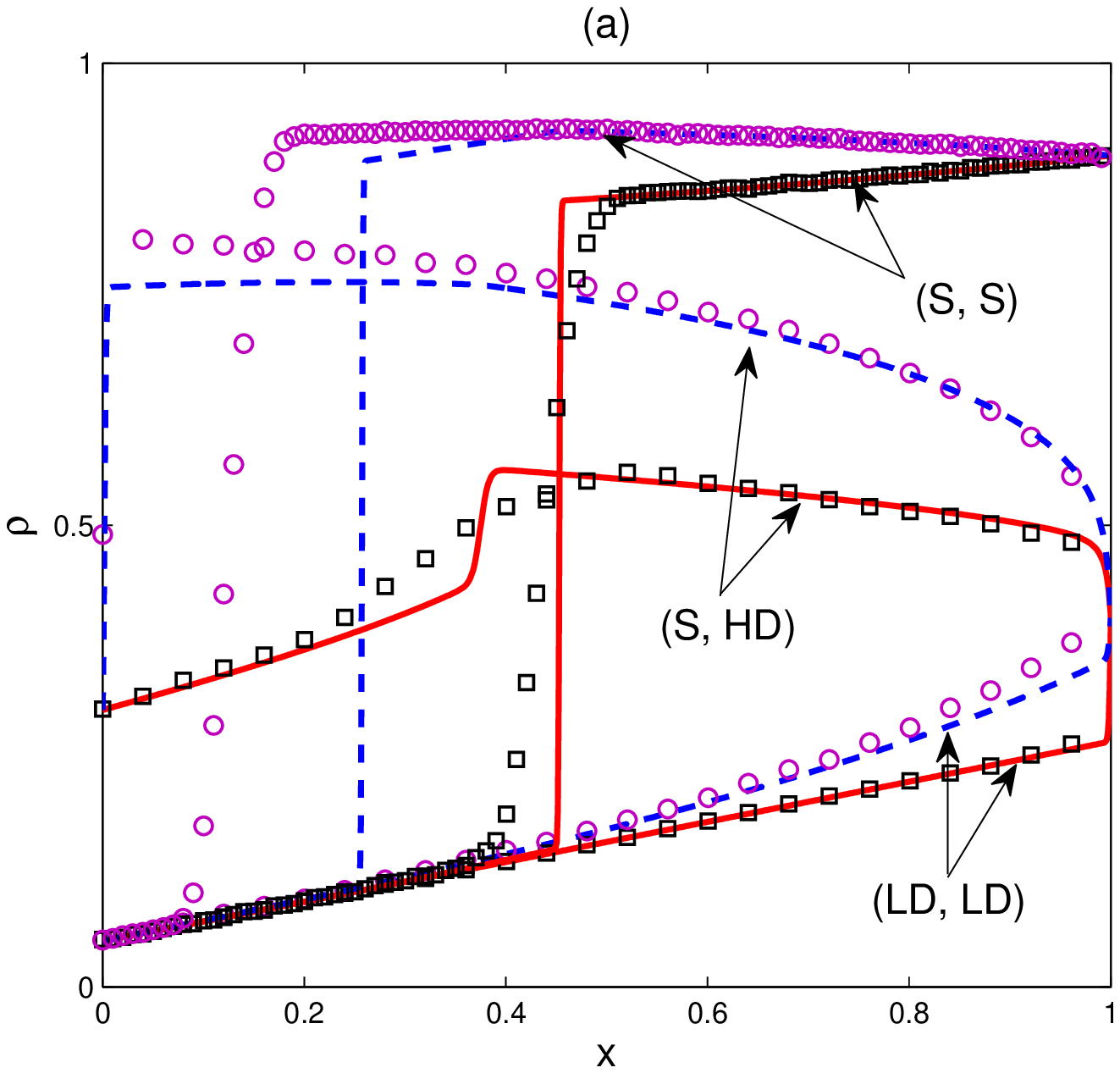}
\includegraphics[width=6.25cm,height=5.5cm]{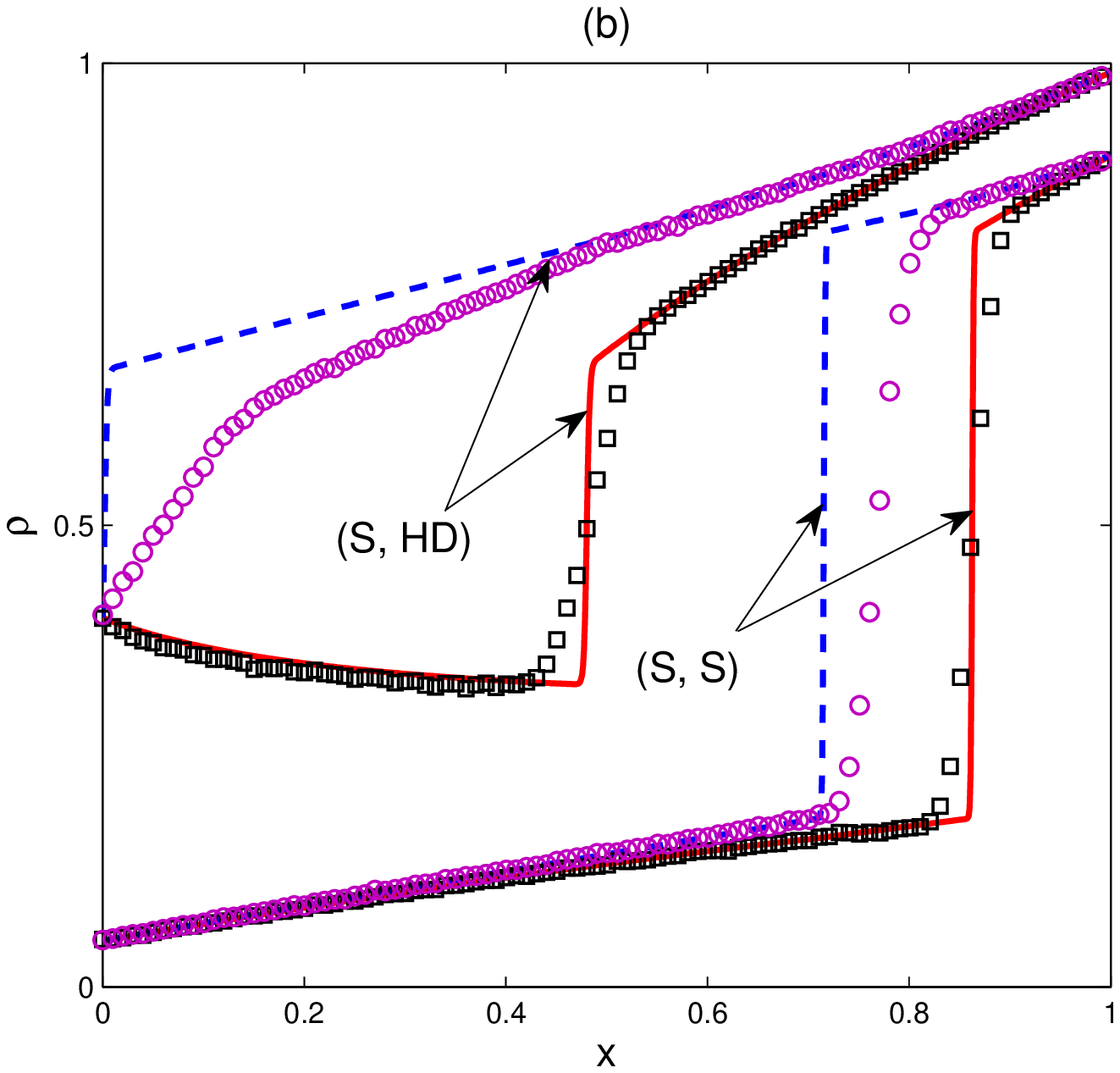}
\caption{(Color online) Density profiles in antisymmetric case (a) (LD, LD) for $\alpha=0.05, \beta = 0.4$; (S, HD) for $\alpha=0.3, \beta = 0.4$ and (S, S) for $\alpha=0.05, \beta = 0.9$ with $\phi=0.5$, (b) (S, S) for $\alpha=0.05, \beta = 0.9$ and (S, HD) for $\alpha=0.4, \beta = 0.99$ with $\phi=-0.5$.}\label{fig:7}
\end{figure}
%%%%%%%%%%%%%%%%%%%%%%%%%%%%%%%%%%%%%%%%%%%%%%%%%%%%%%%%

\begin{figure}
\includegraphics[width=6.25cm,height=5.5cm]{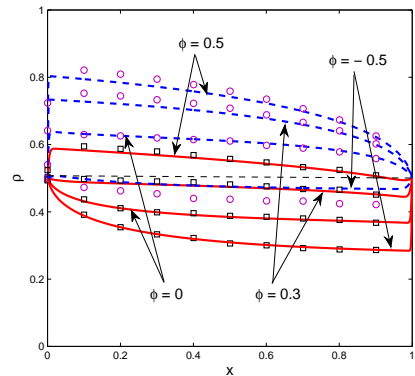}
\caption{(Color online) Phase transition from (LD, LD) to (HD, HD) for $\alpha=0.5, \beta = 0.5$ with respect to an increase in $\phi$.}\label{fig:8}
\end{figure}
%%%%%%%%%%%%%%%%%%%%%%%%%%%%%%%%%%%%%%%%%%%%%%%%%%%%%%%%%%%%%%%%%%%%%%%%%%%%%%%%%%%%%%

\begin{figure}
\includegraphics[width=6.25cm,height=5.5cm]{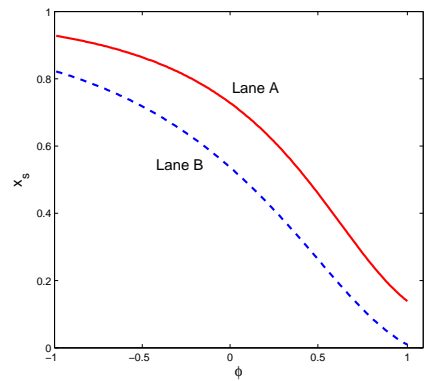}
\caption{(Color online) Effect of $\phi$ on shock position in both the lanes for $\alpha=0.05, \beta = 0.9$.}\label{fig:9}
\end{figure}
%%%%%%%%%%%%%%%%%%%%%%%%%%%%%%%%%%%%%%%%%%%%%%%%%%%%%%%%%%%%%%%%%%%%%%%%%%%%%%%%%%%%%%
\section{Conclusion}
In this work, we study a simple two-lane TASEP model for the transport of molecular motors, which not only interact with each other but also with cellular environment through adsorption/desorption dynamics under a biased lane-changing rule. Based on the recent results on motor protein on microtubules, the mutual interactions are incorporated by the means of modified LK dynamics depending on the configuration of nearest two neighboring sites. To explore the consequences of mutual interactions on the steady-state properties of the system, mean-field equations in the continuum limit are computed and the coupled system is analyzed using a singular perturbation technique. In general, the results of continuum mean-field equations agree reasonably well with Monte-Carlo simulations.

The phase diagrams are obtained for two different cases of mutual interactions. Under the symmetric LK dynamics in which both attachment and detachment rates increase or decrease simultaneously, the topology of the phase diagram remains qualitatively similar to the one obtained in the case of without mutual interaction. The only changes in the structure of the phase diagram are the gradual shifting of the phase boundaries and the shrinkage/expansion of various phases. We have also analyzed the effect of mutual interactions on motion of shocks in the bulk of both lanes, their positions, and heights. For antisymmetric LK dynamics, it is observed that the topology of the phase diagram changes significantly with an increase in attractive/repulsive mutual interaction. The effect of varying mutual interactions on the shocks in both the lanes are different as compared to the one with symmetric LK rates.

The proposed work is an attempt to provide a natural means to qualitatively understand the steady-state properties of a unidirectional transport on a two-channel lattice with mutually interactive LK under fully asymmetric coupling environments. The present study gives some insight not only in understanding complex dynamics of clustering of motor proteins but also towards enhancement of one's insight about various non-equilibrium systems present in nature.

\begin{acknowledgments}
The author gratefully acknowledges the financial support from the Department of Science and Technology (DST), Government of India.
\end{acknowledgments}

%\bibliography{references}

\end{document}